# Offshore oil and gas platform dynamics in the North Sea, Gulf of Mexico, and Persian Gulf: Exploiting the Sentinel-1 archive

Robin Spanier*[a], Thorsten Hoeser[a], John Truckenbrodt[a], Felix Bachofer[a], and Claudia Kuenzer[a,b]

[a]German Remote Sensing Data Center, Earth Observation Center, EOC of the German Aerospace Center, DLR, Wessling, Germany

[b]Institute for Geography and Geology, Department of Remote Sensing, University of Würzburg, Würzburg, Germany

*Corresponding author: robin.spanierdlr.de

**Abtract**

The increasing use of marine spaces by offshore infrastructure, including oil and gas platforms, offshore wind farms, aquaculture, and artificial islands, underscores the need for consistent and scalable monitoring approaches. The development of offshore oil and gas infrastructure has economic, environmental, and regulatory implications. At the same time, maritime areas are difficult to monitor systematically due to their inaccessibility and spatial extent. This study presents an automated approach to the spatiotemporal detection of offshore oil and gas platforms based on freely available Earth observation data. Leveraging Sentinel-1 archive data and deep learning-based object detection, a consistent quarterly time series of platform locations for three of the world's most important production regions: the North Sea, the Gulf of Mexico, and the Persian Gulf, was created for the period 2017-2025. In addition, platform size, water depth, distance to the coast, affiliation with exclusive economic zones, and installation and decommissioning dates were derived. 3,728 offshore platforms were identified in 2025, 356 in the North Sea, 1,641 in the Gulf of Mexico, and 1,731 in the Persian Gulf. While expansion was observed in the Persian Gulf until 2024, the Gulf of Mexico and the North Sea saw a decline in the number of platforms from 2018-2020. At the same time, a pronounced dynamic was apparent. More than 2,700 platforms were installed or relocated to new sites, while a comparable number were decommissioned or relocated. Furthermore, the increasing number of platforms with short lifespans points to a structural change in the offshore sector associated with the growing importance of mobile offshore units such as jack-ups or drillships. The results highlighted the potential of freely available Earth observation data and deep learning methods for consistent, long-term monitoring of marine infrastructure. The derived dataset is publicly available and provides a basis for offshore monitoring, maritime planning, and analyses of the transformation of the offshore energy sector.

## 1. Introduction

Marine areas cover more than 70% of the Earth's surface and are essential for food security, transport, and economic development. At the same time, the expansion of maritime infrastructure has accelerated significantly in recent decades, particularly in the area of energy infrastructure such as offshore oil and gas platforms and offshore wind farms [1, 2]. The spatial distribution and dynamics of these installations have significant ecological, economic, and security impacts [3–6]. Offshore oil and gas platforms are key infrastructure for the exploration, drilling, production, storage, and transport of hydrocarbons [7]. They have different structural configurations that are adapted to water depth and environmental conditions and include fixed platforms as well as temporary or mobile units such as jack-ups, semi-submersibles, or drillships (Figure 1) [8]. This maritime infrastructure contributes significantly to economic development and drives social and technological progress in regions that were previously poor in resources. In 2020, offshore oil and gas production reached a peak with a gross output value of US$988 billion [9]. Despite the global trend toward electrification and the growing share of renewable energies, fossil fuels continue to play a significant role. Global primary energy demand is expected to increase by 23% by 2050, with demand for most primary energy sources continuing to rise [10]. Stakeholders such as governments, regulatory authorities, and operators are seeking to harmonize international practices for the installation and decommissioning of this infrastructure [9, 11]. In many regions, decommissioning is primarily implemented as complete removal, while recent approaches and studies show that alternative decommissioning strategies can also offer environmental and safety-related benefits while reducing economic costs [12, 13]. In this context, former oil and gas fields in the North Sea are increasingly being explored as potential sites for geological $CO_2$ storage, opening up new perspectives for use beyond the traditional life cycle of

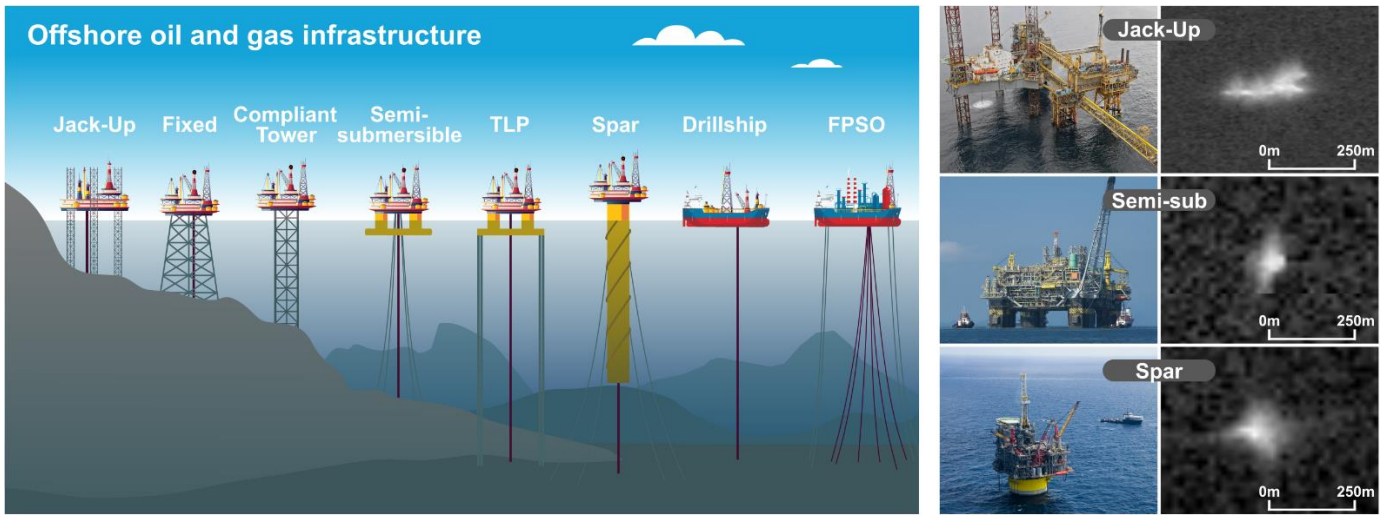

**Figure 1.** Schematic overview of offshore oil and gas infrastructure and their structural and radar image diversity in Sentinel-1 SAR data. The illustration shows different types of installations, categorized by their stability, water depth, and life cycle. Fixed platforms, Spar platforms, and tension-leg platforms (TLPs) are permanent installations often remaining at a single location for decades. In deeper waters, floating production storage and offloading (FPSO) systems and semi-submersibles are used. In contrast, mobile offshore drilling units (MODUs), such as jack-up rigs and drillships, are temporary installations that operate at a single location for only weeks or months. Together, these structures enable the critical phases of the offshore lifecycle, including exploration, drilling, production, storage, and transportation. On the right are examples of aerial images and the corresponding Sentinel-1 signatures for three types, which illustrate the high variability in shape, extent, and backscatter signature. Figure adapted from [14]; Aerial image of the semi-submersible retrieved from commons.wikimedia.org, credit to [15].

offshore infrastructure [16–19]. Effective management of this dynamic development of the maritime space requires robust, spatially consistent, and high-resolution information on offshore oil and gas platforms, including spatial attributes and lifecycle information [3, 6, 9]. Such data is key to maritime awareness, environmental management, risk assessment, regulatory planning, and the evaluation of transition pathways in the energy sector [9, 11, 20]. However, monitoring human activities at sea remains challenging due to the inaccessibility of maritime spaces and the large scale of the areas to be observed. Although reporting requirements and monitoring approaches exist, such as via Automatic Identification System (AIS) or regulatory instruments such as the Environmental Impact Assessment (EIA) Directive in the EU [21, 22], these are regionally limited, inconsistently enforced, and often incomplete due to security and confidentiality concerns [23–26]. In response to these challenges, scalable, robust methods for the automated detection and characterization of offshore oil and gas platforms and for evaluating their spatiotemporal development are essential. Time series analyses, supplemented by spatial attributes and life cycle information, open up new possibilities for the consistent quantification of infrastructure developments.

Over the last decades, large amounts of satellite images have been systematically collected and archived, creating new opportunities for monitoring environmental changes and infrastructure [27–29]. Satellite remote sensing enables regular, spatially consistent observation without direct access and is particularly relevant for regions that are difficult to access, such as the high seas [30, 31]. Global coverage, high temporal frequency, and the availability of long-term archives allow large-scale developments to be analyzed as consistent time series. A significant share of this data comes from public programs such as Landsat and Copernicus and is freely available, which greatly facilitates transparent monitoring studies.

The detection of persistent marine infrastructure from Earth observation data is used in several studies to describe the development of maritime energy infrastructure and to better understand the impact of human activities at sea. Previous work on the automated detection of offshore targets is often based on rule-based methods, including constant false alarm rate (CFAR) detectors or threshold-based methods in combination with geometric or multiscale filters [32–34]. Wong et al. [24] combined CFAR-based detection with difference-of-Gaussians and downstream filtering to derive offshore infrastructure from Sentinel-1 data. Concurrently, optical and multispectral time series were also used to detect marine infrastructure [26, 35, 36]. However, such optical approaches are limited for consistent offshore monitoring due to atmospheric disturbances such as clouds and haze. Synthetic aperture radar (SAR), therefore, represents the key alternative, as it operates independently of weather conditions and offshore objects typically appear as bright backscatter signatures against the darker sea background (Figure 1). Offshore platforms form characteristic backscatter clusters, whose observed extent is larger than their physical structure. This is due to signal geometric effects such as layover and double and multiple scatterings between vertical platform structures and the water surface [33, 37]. These distinct and comparatively stable backscatter signatures form the basis for their multitemporal identification in SAR time series. A widely used processing step is the creation of multitemporal composites to suppress mobile targets such as ships and highlight static signatures of infrastructure, both purely SAR-based [24, 38] and in combination with optical data [6, 39]. On this basis, large-scale applications have increasingly been implemented. Zhang et al. [40] processed Sentinel-1 archives globally and derived offshore wind turbine locations using a morphological approach and multiple thresholds. With the constantly increasing availability of large EO data archives and powerful data-driven methods, the focus of research has shifted toward deep learning. However, the limited availability of high-quality annotated training data and the scaling of models beyond locally optimized case studies to large-scale and heterogeneous marine regions remain key challenges [14]. Hoeser et al. (2022) provided DeepOWT, a global dataset on offshore wind energy infrastructure derived from Sentinel-1 median composites using a cascade of two convolutional neural networks (CNNs) optimized on synthetic training data [29, 38, 41].

Collectively, these studies demonstrate the potential of SAR-based time series for large-scale detection of marine energy infrastructure. For offshore oil and gas platforms in particular, integrating time series information with spatial attributes enables consistent analysis of the dynamics of maritime energy infrastructure.

The motivation for this study stems from the critical need for automated spatiotemporal monitoring of offshore oil and gas infrastructure using Earth observation data. Building on the Sentinel-1 detection model developed and validated in Spanier and Kuenzer (2026) [42], offshore oil and gas platforms in three key regions (North Sea, Persian Gulf, Gulf of Mexico) are systematically analyzed using time series, from which quarterly products are derived. The core contributions of this work are threefold:

- A consistent quarterly inventory of offshore oil and gas platforms from 2017 to 2025 across the three key production regions, including lifecycle (first appearance and removal) and key spatial attributes (exclusive economic zones, coastal distance, water depth, and platform size).
- The provision of the identified offshore oil and gas platforms, including temporal information and derived characteristics, as an open dataset in GeoParquet format.
- A scalable workflow for deriving spatiotemporal platform statistics, applicable to the entire Sentinel-1 archive and extendable to global applications.

## 2. Materials and methods

The methodology used in this study is based on a modular workflow for the consistent derivation of a quarterly time series of offshore oil and gas platforms, including life cycle information and key spatial characteristics. Figure 2 summarizes the process chain, from the acquisition and preprocessing of Sentinel-1 data to inference with a pre-trained detection model to postprocessing, temporal linking, and spatial feature enrichment. The individual modules are described in detail in the following subsections.

### 2.1. Study regions

The study focused on three major offshore oil and gas production regions: the North Sea (NS), the Persian Gulf (PG),

and the Gulf of Mexico (GoM) (Figure 2). The boundaries of the regions followed the official definitions of the International Hydrographic Organization [43]. The three regions differ significantly in terms of geographical conditions and offshore infrastructure.

The North Sea (NS) is a semi-enclosed shelf sea in Northern Europe with an area of approximately 750,000 km², shared by several highly industrialized coastal states (including the United Kingdom, Norway, Denmark, Germany, the Netherlands, Belgium, and France). With an average water depth of approx. 80-90 m it is relatively flat overall, but deepens towards the north to the Norwegian Trench, reaching depths of up to ~700-800 m [44, 45]. In the wake of the energy policy transformation and stricter climate goals, the region is also undergoing structural change, with traditional oil and gas production increasingly being supplemented by the expansion of offshore wind energy and new uses for existing infrastructure [9, 46]. This shift is also reflected in a projected decline of 0.8% per year in European oil demand by 2030 [46].

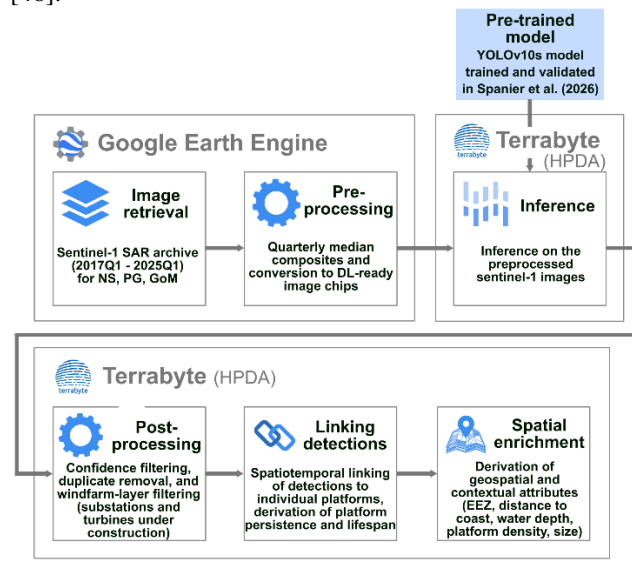

**Figure 2.** Workflow of the spatiotemporal development analysis of offshore oil- and gas platforms in the North Sea, Gulf of Mexico and Persian Gulf. The workflow includes the key processing steps from S1 data acquisition and preprocessing, model training with real and synthetic training data, inference and post-processing, spatial enrichment and lifespan analysis.

The Persian Gulf is a shallow, semi-enclosed basin measuring just under 1,000 km in length and covering an area of approximately 240,000 km², bordered by the United Arab Emirates, Saudi Arabia, Qatar, Bahrain, Kuwait, Iraq, Iran, and Oman. The average water depth is around 36-40 m, with extensive coastal areas less than 20 m deep; deeper sections occur mainly along the Iranian coast and towards the Strait of Hormuz. The combination of shallow depths, very large oil and gas reserves, and the global strategic importance of the Strait of Hormuz has led to the development of a particularly dense and geopolitically relevant offshore infrastructure [47, 48]. Given that around 20% of global oil consumption passes through the Strait of Hormuz every day [49] and that political tensions in the region, most recently in early 2026, have a direct impact on energy prices and security of supply, this bottleneck plays a key role in global energy politics and security. The projected oil demand in the Middle East for the period 2024-2030 is expected to remain stable, further highlighting the region's enduring importance for global energy security and supply stability [46].

The Gulf of Mexico (GoM), on contrast, forms a significantly larger, semi-enclosed deepwater basin (approx. 1,500,000 km², ) with an average water depth of about 1,615 m, resulting in pronounced deepwater conditions, particularly in the central basin. Politically, the region is bordered by the exclusive economic zones of the US, Mexico, and Cuba, with offshore oil and gas infrastructure concentrated primarily in the northern (US) and southwestern to southern (Mexican) Gulf marginal zones [50]. At the same time, the Gulf of Mexico is one of the global hotspots for offshore platform density, with the long-term decommissioning and removal of aging structures playing a key role [9]. Within this context, the International Energy Agency (IEA) forecasts a slight decline of 0.4% per year in oil demand for North America by 2030 [46].

### 2.2. Image retrieval and preprocessing

SAR data from the Copernicus Sentinel-1 mission was used, due to its proven effectivity for detecting metallic offshore targets, availability regardless of cloud cover, and a long-standing, freely accessible data archive [51]. The S1 constellation consists of four C-band SAR satellites. S1A (launched in 2014) and S1B (2016) are in sun-synchronous orbits with a phase shift of 180° and offer a combined repeat rate of 6 days. Following the failure of S1B in 2021, S1C was launched in 2024 to maintain continuity. In addition, the coverage frequency varies regionally depending on orbital overlap and mission priorities, resulting in denser coverage over Europe [52] (Figure 2).

We used the Sentinel-1 Level-1 GRD product provided in Google Earth Engine [53]. Processing includes orbit data correction, removal of edge artefacts and thermal noise, radiometric calibration, and terrain correction, and provides the backscatter coefficient σ° in decibels (dB). For the analysis, VH polarization data in Interferometric Wide (IW) Swath mode with a pixel spacing of approximately 10 m was used. The Sentinel-1 archive was queried for all scenes covering the study areas and acquired between 2016Q1 and 2025Q1. Due to insufficient coverage in 2016 (Figure 2b), the time series analysis begins with 2017Q1.

To enable efficient data processing and export from GEE, a grid with a step size of 1.8° was generated, dividing the study areas into UTM-WGS84 tiles (Figure 3a). Within GEE, scenes were stacked per quarter and grid tile and aggregated into median composites to suppress mobile targets such as ships, while offshore structures operating stationary over longer periods of time remain in the composite. To reduce the data volume, the 16-bit backscatter values were converted to 8-bit integers by capping the value range from -40 dB to 0 dB and mapping it linearly to 0-255. The resulting scaled 8-bit median composites were then exported from GEE. The subsequent chipping represents the first processing step that was carried out on the on-premises infrastructure. The exported composites were split into 640×640-pixel chips with 20% overlap to generate model-compatible inputs and reduce edge artifacts. A total of 4,103 quarterly median composite tiles was generated for all study areas and 33 quarters between 2017Q1 and 2025Q1 (Table 1).

**Table 1**. Summary of Sentinel-1 satellite imagery downloaded and used for each study region and 33 quarters (2017Q1-2025Q1). The table demonstrates the significant regional variation in data availability, particularly with regard to acquisition density. Asterisk (*) indicates the approximate size based of the boundaries defined by the International Hydrographic Organisation (IHO) [43].imagery downloaded for each study region and 2023Q4.

| Regions of interest | approx. area (km²) * | median comp. tiles |
|---|---|---|
| North Sea | 570,000 | 1,419 |
| Persian Gulf | 250,000 | 594 |
| Gulf of Mexico | 1,340,000 | 2,079 |
| **Sum ∑** | | 4,103 |

### 2.3. Model

For the automatic detection of offshore oil and gas platforms, a deep learning-based object detection YOLOv10 model [54] was used. The model described in [42], which was trained and validated on independent regions, was adopted and applied to a multi-temporal Sentinel-1 time series as part of this study. The aim was to utilize the generalization capability demonstrated in [42] for a spatiotemporal analysis of platform dynamics.

The training dataset for that model was based on Sentinel-1 data from quarterly median composites. The ground truth was created manually and supplemented with optical references and external data. Two platform classes (single platform and platform cluster) and wind turbines were defined as an off-target class. To improve class balance, the dataset was supplemented with synthetic image-label pairs. The training was based exclusively on data from the South China Sea, Caspian Sea, Gulf of Guinea, and the Brazilian coast, while the North Sea, Persian Gulf, and Gulf of Mexico were completely withheld as an independent test dataset. This geographic region holdout design allowed for an explicit assessment of transferability to previously unseen marine regions. In addition, it was ensured that image chips from the same Sentinel-1 tile were not scattered across training, validation, and test splits. The model provides bounding boxes with class probability for each detection. Evaluation on the independent test dataset resulted in a precision of 0.91, a recall of 0.89, and an F1 score of 0.90 for the combined platform class at a confidence threshold of 0.5 and intersection-over-union (IoU) of 0.3 [42].

### 2.4. Inference and postprocessing

The trained YOLOv10 artifacts were transferred to the High-Performance Data Analytics (HPDA) platform "terrabyte" of the German Aerospace Center (DLR) and the Leibniz Supercomputing Centre (LRZ). The platform provides GPU-based computing resources (NVIDIA A100) and storage capacity for processing large Earth observation datasets. Inference was performed for all pre-processed image chips for each quarter in 2017 to 2025.

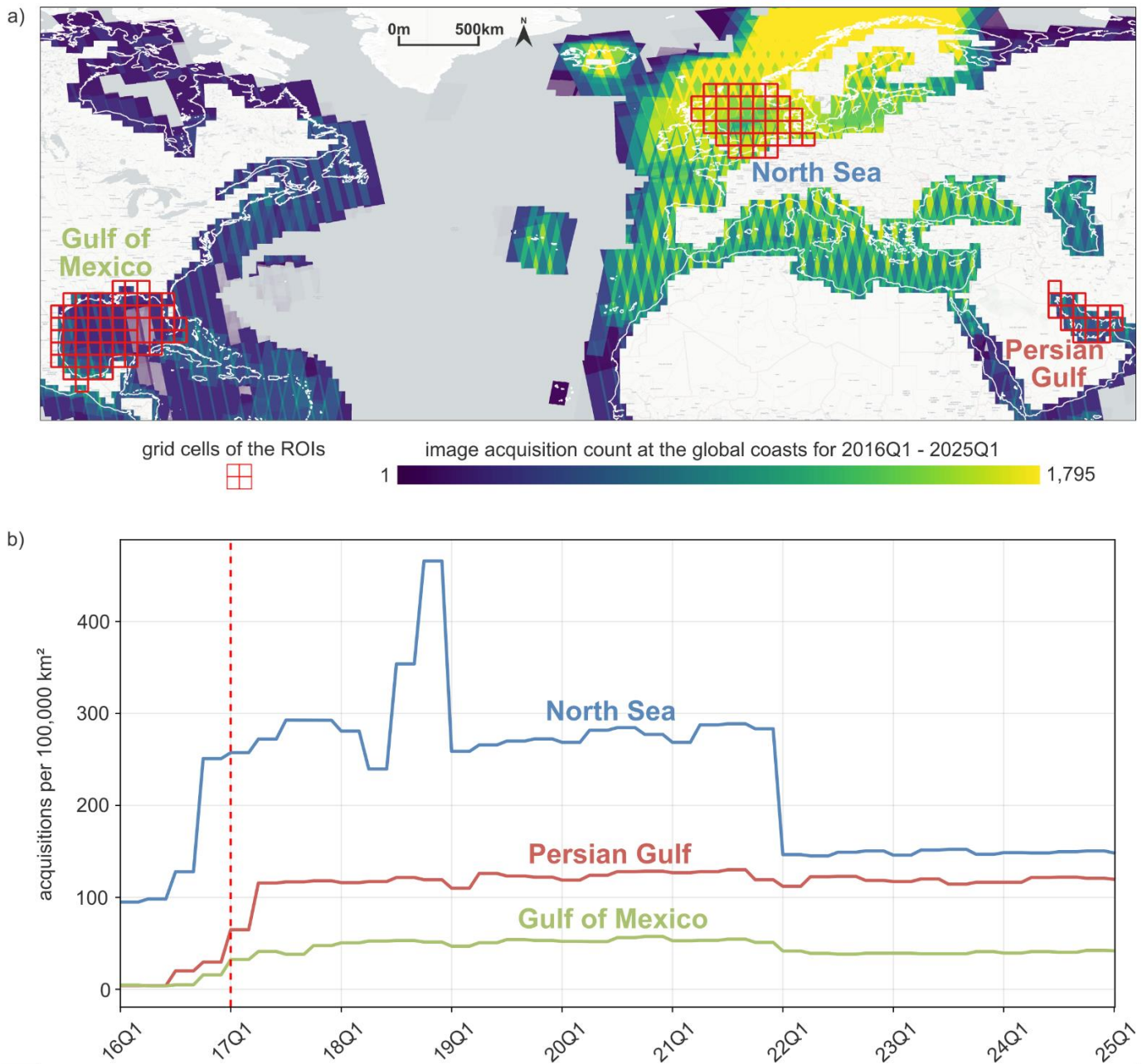

**Figure 3.** Overview of the three study regions North Sea, the Gulf of Mexico, and the Persian Gulf and the Sentinel-1 acquisition density between 2016Q1 and 2025Q1. a) displays the spatial distribution and image acquisition count of Sentinel-1 SAR scenes that overlap with the global coasts between 2016Q1 and 2025Q1. It demonstrates the significant regional variation in acquisition density (higher density in Europe, lower density in America). Study region boundaries follow the International Hydrographic Organization (IHO) [43] definitions. b) shows the available Sentinel-1 acquisitions per 100,000 km² for each region, illustrating temporal variability and the lower data availability in 2016.

The predictions were exported as text files with detection metadata and georeferencing.

After inference, all detections from all image chips in a quarter were merged. Bounding boxes were reprojected to EPSG:4326 (WGS84) to enable spatial operations, and each detection was assigned a unique identifier. Predictions below the defined confidence threshold of 0.4 were discarded. To reduce noise, detections with pixel values entirely below 150 (≈ -16.5 dB) were removed. This threshold was derived from the analysis of the radar backscatter characteristics of verified platforms in the 8-bit S1 composites. To reduce duplicates due to tile overlaps, overlapping detections were grouped using an IoU threshold of 0.2. For each group, the most reliable prediction was selected based on class consensus and confidence value. Since the underlying model also detects offshore wind farm substations and wind turbine foundations under construction, these were removed from the dataset using an external offshore wind farm layer [55]. Manual filtering or removal was deliberately not performed to ensure the transparency, reproducibility and scalability of the automated workflow. The cleaned detection set was exported as GeoJSON for further spatial analysis.

## 2.5. Derivation of individual platforms and lifespan

To derive individual offshore platforms from quarterly detections of the same physical structure, a spatiotemporal consolidation approach was applied. Spatially overlapping bounding boxes were merged into clusters across all quarters using IoU (IoU ≥ 0.1). Each cluster group represents a single physical platform with a unique platform ID. For each platform, the associated observed quarters were sorted chronologically. The lifetime of a platform was defined as the time span between its first and last detected occurrence. For this interval, a continuous presence of the platform was taken as a basis, in line with the physical and economic characteristics of offshore oil and gas production. Offshore platforms are linked to the continuous exploitation of fixed hydrocarbon deposits, whose operation extends over a continuous period until depletion and is not characterized by repeated short-term installation and decommissioning cycles. The chosen consolidation approach also reduced artifacts that can arise from temporary fluctuations in satellite coverage or detection performance and could otherwise lead to artificial short-term changes in the number of platforms. The resulting dataset of individual platforms formed the basis for the subsequent analyses.

## 2.6. Spatial enrichment

For further analysis, additional spatial attributes were derived and added to the individual offshore platforms. Political and physical-geographical characteristics as well as structural properties of the platform locations were factored in. The national affiliation and regional classification of the platforms were determined by spatially linking the platform centers with exclusive economic zones (EEZs) and regions.

Spatial queries with external geodata were performed to determine physical-geographical location parameters. The minimum distance to the coastline was calculated as the geodetic distance between the platform center and the nearest coastline by using a global coastline dataset (10 m) including major island structures (provided by Natural Earth).

The water depth at the platform location was determined by point query of a GEBCO global bathymetry raster [56] at the platform center. To minimize NoData pixel artifacts, a local 3×3 neighborhood window was also taken into account where necessary. Negative values correspond to sea depth below sea level. These attributes enable consistent quantification of platform distribution along bathymetric gradients and different coastal distances.

The size estimation is based on the detected bounding box areas, since detection was implemented as an object detection task on SAR backscatter signatures and not instance segmentation of the real platform geometries, e.g., on optical satellite images. Due to the known larger sized backscatter effects of metallic offshore structures in the SAR signal, these values do not represent exact physical platform dimensions, but they do allow for consistent relative comparability between regions and time periods.

## 2.7. Offshore platform dataset (OPD)

Based on the curated platform data described above, the Offshore Platform Dataset (OPD) was created and made publicly available. The dataset is provided in three complementary product files. The first file includes all offshore oil and gas platforms detected between 2017Q1 and 2025Q1 in the three study regions. For each platform, the first and last observed quarter, the distance to the coast, the water depth, the bounding box area, the national EEZ affiliation, the region (NS, PG, GoM), and the geographical position (latitude/longitude) are provided. The second file represents platforms for each quarter between 2017Q1 and 2025Q1 and their derived attributes. The third file represents a snapshot for 2025Q1 and contains only platforms detected in that quarter and their derived attributes, providing an up-to-date inventory of offshore oil and gas infrastructure in the three study regions. The dataset is published under the name Offshore Platform Dataset (OPD v1.0.0) in GeoParquet format (.parquet) and is freely accessible via Zenodo (https://doi.org/10.5281/zenodo.19046824). The dataset supports the systematic capture and updating of existing infrastructure inventories and provides a reliable data basis for decision-makers and for further analysis, for example in the areas of environmental monitoring, infrastructure development, and spatial planning.

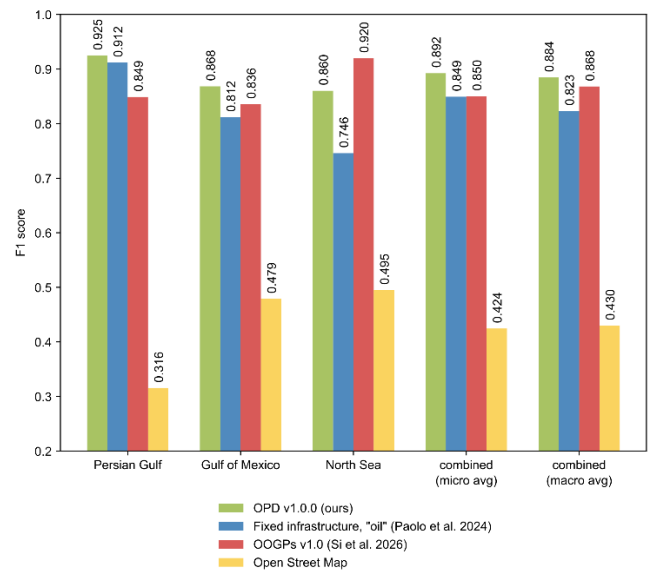

**Figure 4.** Comparison of offshore oil and gas platform detection performance of several freely available datasets for the 2023 ground truth test dataset of the three study areas. Across all three regions, OPD v1.0.0 achieved the highest overall performance among the freely available datasets with a macro-weighted F1 score of 0.884, exceeding Paolo et al. (0.823) [39], OOGPs v1.0 (0.868) [57], and OpenStreetMap (0.430). Regionally, OPD v1.0.0 attained the highest F1 scores in the Persian Gulf (0.925) and the Gulf of Mexico (0.868). In the North Sea (0.860), performance remained above Paolo et al. and OpenStreetMap, while OOGPs v1.0 reached a higher F1 score (0.920). Overall, the results indicate consistent cross-regional performance and robust coverage of offshore oil and gas infrastructure across heterogeneous offshore regions.

We compared OPD v1.0.0 with other datasets on offshore oil and gas infrastructure covering the three study regions (North Sea, Persian Gulf, Gulf of Mexico) and that are freely available without access restrictions. The evaluation was performed per region and aggregated (macro and micro averaged) using standard accuracy metrics (precision, recall, F1 score) and the reference ground truth dataset from Spanier et al. 2026 [42]. Figure 4 shows the F1 scores.

## 3. Results

### 3.1. Spatiotemporal development of offshore oil and gas platforms

second Since 2017, the Gulf of Mexico has had the highest number of offshore platforms, followed by the Persian Gulf and the North Sea (Figure 5a). In all three regions, an initial increase in platform numbers was observed between 2017 and around 2019, coinciding with a phase of high global offshore investment activity and new offshore projects in the US Gulf of Mexico [9]. The respective maxima were reached in the period 2018-2020. After that, developments diverged significantly. While further continuous growth was evident in the Persian Gulf until 2024, the Gulf of Mexico and North Sea showed a steady decline from around 2019/2020 onwards. The sharp decline in 2020 coincided with the onset of the COVID-19 pandemic, which, according to the OECD (2025) [9], represented a significant external shock to large parts of the ocean economy, including offshore oil and gas activities. Following the decline in the Gulf of Mexico and the simultaneous increase in the Persian Gulf, the platform numbers for both regions converged over time. In 2024Q4, the number of platforms in the Persian Gulf surpassed that of the Gulf of Mexico for the first time. This shifted the ranking of the regions at the end of the time series. At the same time, however, a decline in platform numbers was also recorded in the Persian Gulf from 2024 onwards, following a period of increasing geopolitical tensions in the Middle East since 2023. The North Sea remains below the other two regions in platform numbers throughout the entire period and showed a continuous downward trend since 2019 until 2025Q1. This development is consistent with efficiency gains and energy policy transformation processes in the context of long-term net-zero strategies in European countries [9].

At the level of national affiliations (Figure 5b), it is evident that regional trends are significantly influenced by individual countries. In the Gulf of Mexico, the detected platforms account for the United States and Mexico. The Gulf of Mexico curve is almost identical to that of the US, indicating its dominant share within the region. With more than 1,500 platforms over the entire study period, the US recorded the highest absolute values of all the countries considered. Mexico, on the other hand, showed a comparatively stable, slightly upward trend in the range of around 200-250 platforms. In the Persian Gulf, the trend was primarily determined by Saudi Arabia and the United Arab Emirates. While the United Arab Emirates and the other neighboring countries showed relatively constant platform numbers, the significant growth in the Persian Gulf between 2017 and 2024 was mainly driven by the increase in Saudi Arabia.

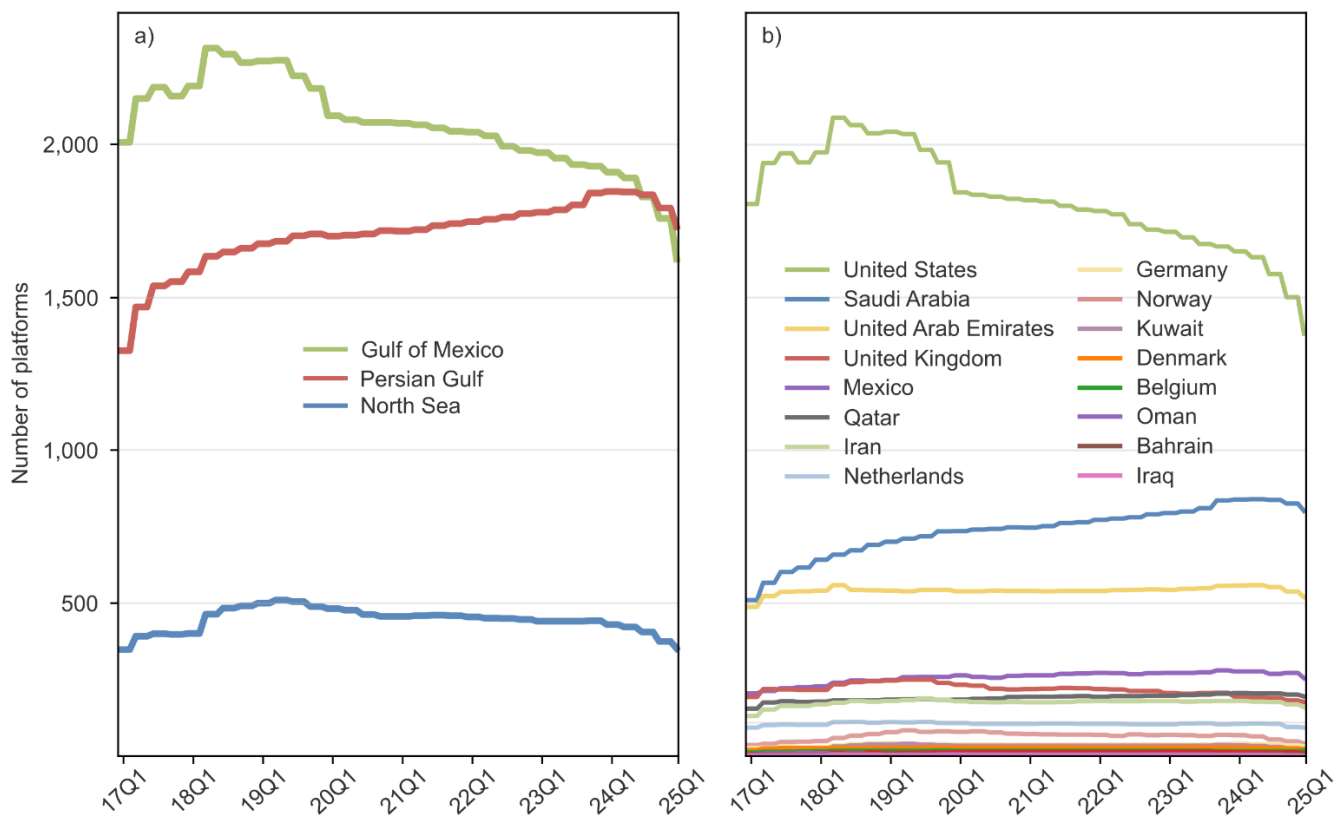

**Figure 5.** Offshore oil and gas development in 2017-2025 in the three study regions of the North Sea, Persian Gulf, and Gulf of Mexico (a) as well as in each country's EEZ (b). The time series represent the quarterly number of offshore oil and gas platforms detected.

The dynamics of the regional aggregate curve thus essentially reflected developments in Saudi Arabia. In the North Sea, the United Kingdom, the Netherlands, and Norway made the largest contributions to platform counts. The Netherlands showed a largely stable number of platforms over the entire period. The United Kingdom and Norway recorded a temporary increase between 2018 and 2019, followed by a gradual decline until 2025Q1.

The analysis of coastal distance (Figure 6) reveals clear regional differences between the study areas, whereas no substantial temporal variation was observed during the study period from 2017Q1 to 2025Q1. In the Persian Gulf, only comparatively short coastal distances (median = 53 km) are possible due to the limited width of the basin. In the Gulf of Mexico, the platforms were located at a median distance of 24 km from the coast and were concentrated along the extensive continental shelf, but locally extended to the shelf edge and towards the central deep-sea zone. In the North Sea, on the other hand, the platforms were spread over large areas of the central basin and regularly reached distances of over 200 km, in some cases close to 300 km (median = 92 km). The aggregated distributions confirm this significantly greater offshore extent of the North Sea compared to the other two regions.

Water depth also shows clear regional patterns that are visible in the aggregated distributions of the three regions. In all three regions, the platforms were predominantly concentrated in shallow waters, with median values ranging from approximately 18 m (GoM), 27 m (PG), and 38 m (NS). Around 90% of all detected platform were located in water depths of no more than 100 m. The deepest waters in the North Sea (up to around -350 m) were found at the installations of the Norwegian Troll field in the northern North Sea basin west of Bergen. In the central deep-sea basin of the Gulf of Mexico, with water depths of over 3,000 m, only a few platform locations were recorded in the data. The maximum detected water depths were attributable to deepwater installations such as the Perdido platform (2,450 m) and the Turritella FPSO in the Stones field (over 2,900 m). Over time, there were no systematic shifts towards greater water depths (see Supplementary Figure 1). The distribution of locations thus primarily reflects the technical and economic usability of marine areas and not solely their geometric proximity to the coast.

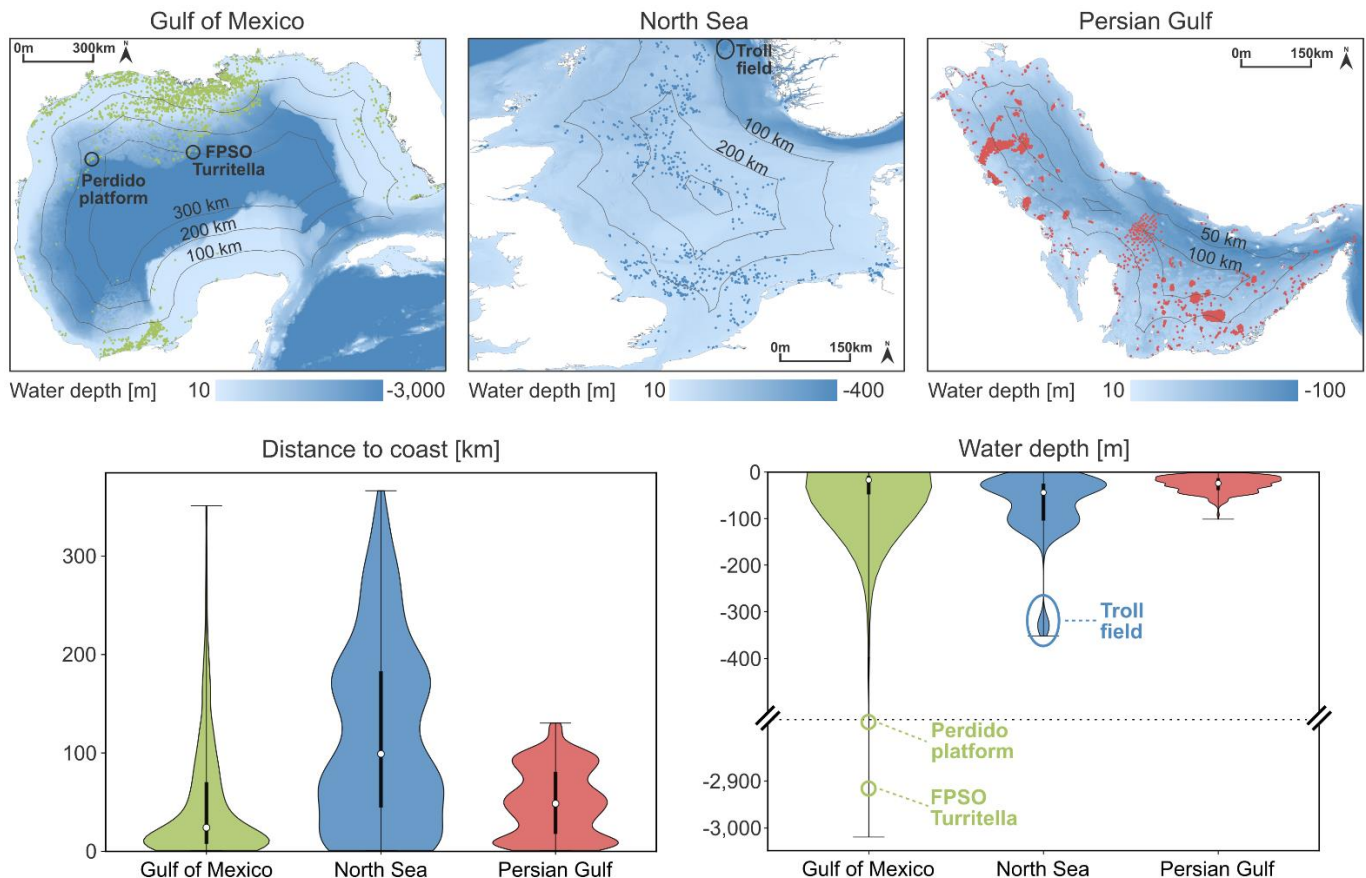

**Figure 6.** Regional distribution of offshore oil and gas platforms in relation to water depth and distance from the coast aggregated for the period 2017Q1-2025Q1. Top: Overview maps of the three study regions with all platforms detected during the study period. A coastal distance buffer and a bathymetry grid are displayed. Note the different value ranges of the water depth scales between the regions. Bottom: Aggregated distributions of distance to the coast and water depth for each study region as violin plots. Each violin represents the distribution of all detected platform locations over the entire study period. Note the gap in the y-axis of the water depth chart, which was added to cover all platforms, including the outliers in deep waters in the Gulf of Mexico.

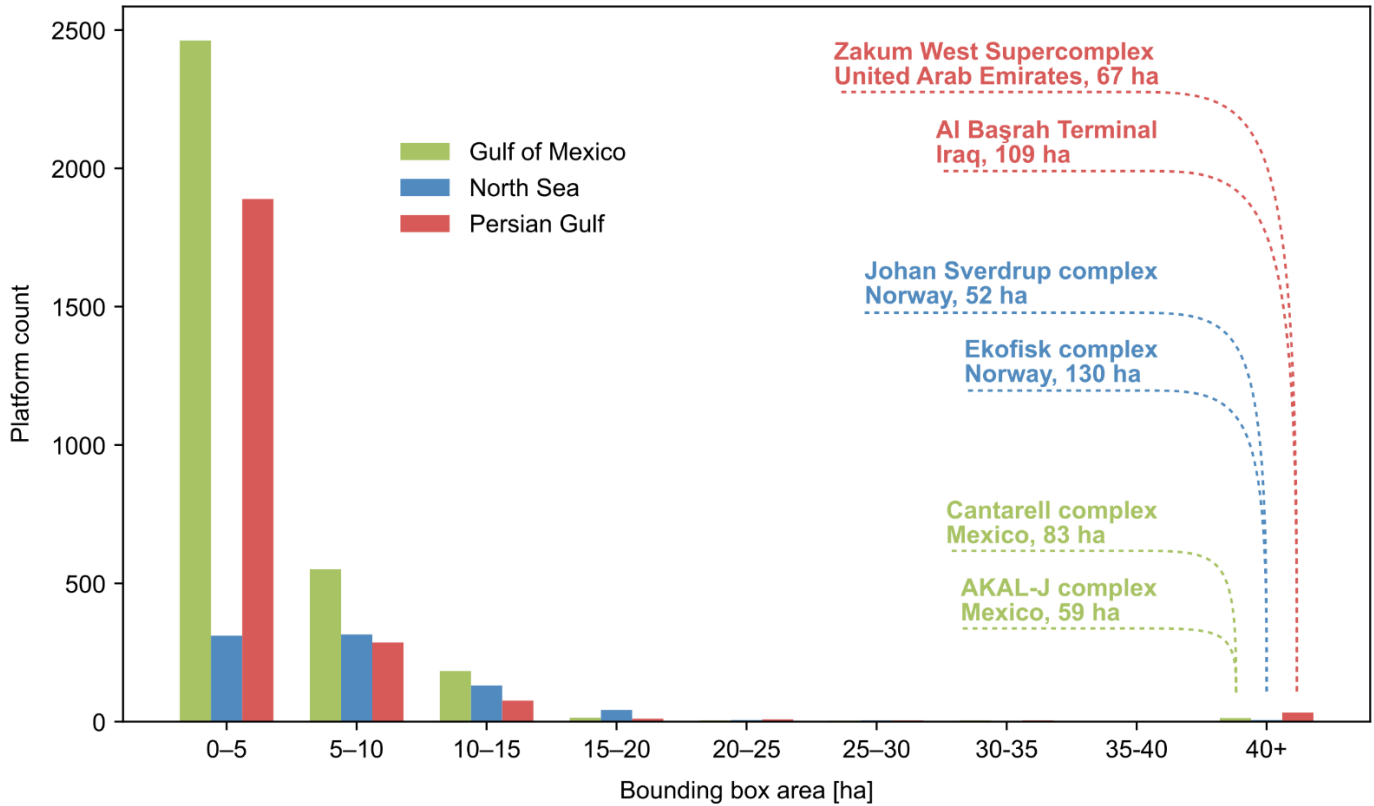

**Figure 7.** Distribution of platform size based on the detected bounding box areas in the period 2017Q1-2025Q1. The distributions shown illustrate both the various typical platform sizes and the presence of very large platform complexes, particularly in the Persian Gulf.

Consistent spatial aggregation patterns can be observed within suitable depth zones.

Platform size (Figure 7) shows clear regional differences, but showed no significant changes over the entire study period. No structural shifts toward larger or smaller platform types were observed. Platforms in the North Sea were larger on average (5.33 ha) than those in the Gulf of Mexico (2.45 ha) and the Persian Gulf (2.04 ha). While 87.2% of platforms in the Persian Gulf and 81.9% in the Gulf of Mexico were smaller than 5 ha, this applied to only 47.9% in the North Sea, where 31.9% were 5-10 ha and 20.3% with areas of over 10 ha.

The majority of all structures in all three study regions are smaller than 20 ha. Platforms ranging from approximately 20 to 40 ha are comparatively rare, while very large platform structures do not appear again until bounding box sizes of 40 ha or more. This distribution highlights the infrastructural diversity of offshore oil and gas infrastructure types as described in Chapter 1. Such large-scale platform complexes are particularly prevalent in the Persian Gulf, followed by the Gulf of Mexico. This shows that, while the platforms installed in the North Sea are larger on average than those in other regions, the largest individual structures or platform complexes are primarily found in the Persian Gulf.

**3.2. Platform lifecycle dynamics**

Figure 8 visualizes lifespan dynamics of each individual offshore platform location between 2017Q1-2025Q1. Each individual platform is represented by a horizontal bar. The Gantt charts show whether platforms existed continuously, were newly added within the observation period, or continued to persist beyond it.

The majority of platforms were continuously present in all three regions throughout the entire study period or extended beyond its bounds. In the Persian Gulf, this applied to 79% of platforms, in the Gulf of Mexico to 71%, and in the North Sea to 55%. These installations either existed before 2017 or were still active in 2025. This illustrates that the observation period from 2017 to 2025 only represents a snapshot of infrastructure development compared to the long-term use of stationary production platforms, many of which operate for decades. A high proportion of permanently detected platforms therefore primarily reflects the long-term infrastructural stability of established offshore fields. At the same time, however, the presence of temporary offshore units is also visible, including, for example, jack-up platforms, FPSOs, or drillships, which are only stationary operating until relocation. Platforms with medium (5-8 years) or short (<5 years) presence durations accounted for 45% of locations in the North Sea (38% and 7% respectively), 28% in the Gulf of Mexico (23% and 5% respectively) and 22% in the Persian Gulf (18% and 4% respectively). This clearly highlights regional differences in infrastructure persistence: the North Sea has the highest proportion of temporary installations, while long-term installations clearly dominated in the Persian Gulf. Furthermore, in all three regions, newly installed platforms tended to have a shorter presence than older installations. This consistent trend across all basins points to increasing operational flexibility and a structural increase in temporary, mobile offshore units. This dynamic is also reflected in the high turnover rate of offshore infrastructure the three regions. Between 2017 and 2025, a total of 2,763 platforms were newly installed or relocated to new sites, while 2,718 platforms were decommissioned or relocated.

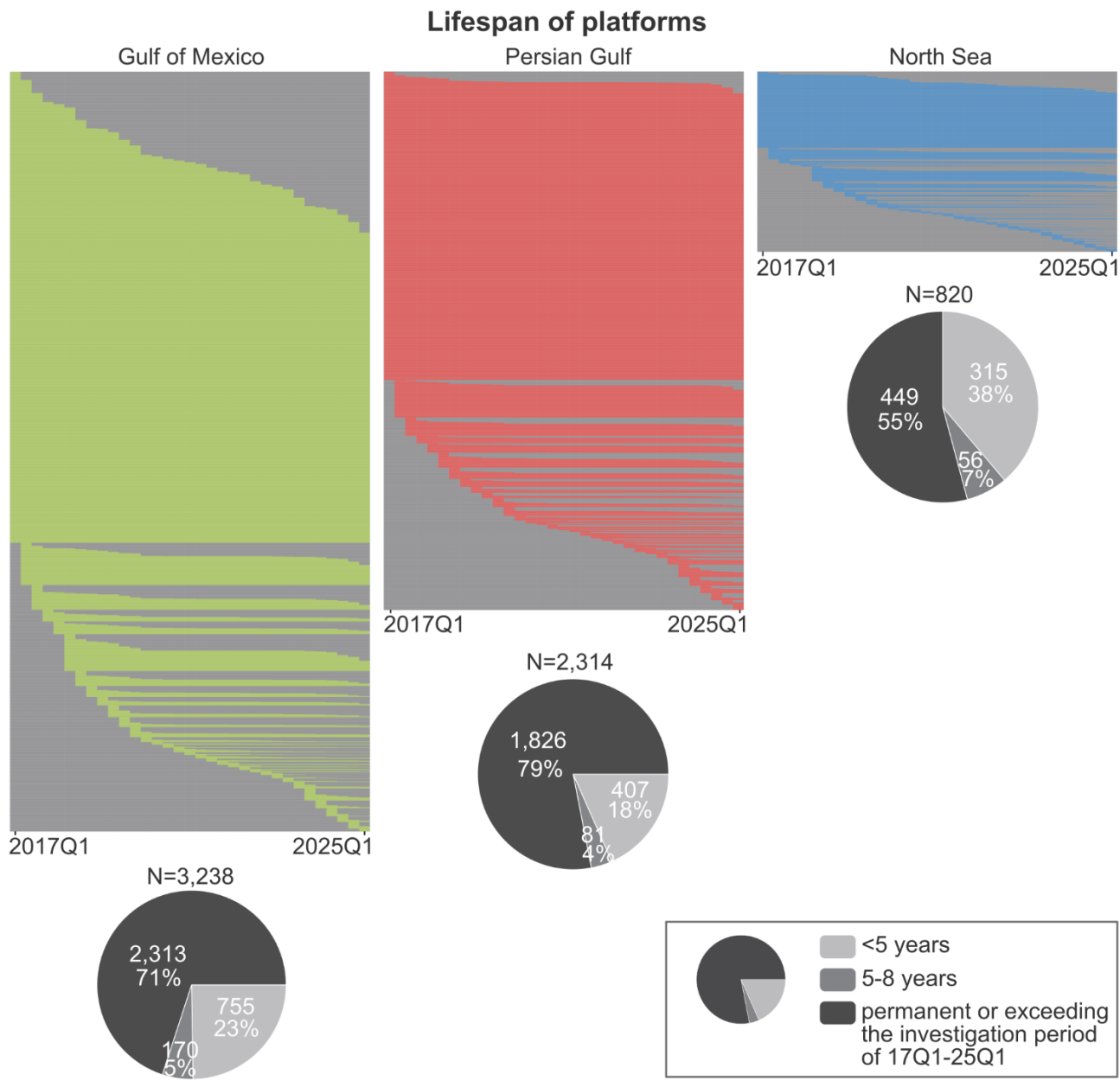

**Figure 8.** Lifespan dynamics of offshore oil and gas platforms in the period 2017Q1-2025Q1. The upper panels show presence Gantt charts at the individual platform level (one bar per platform). Colors marks periods during which a platform was present, gray marks periods before deployment or after removal or relocation. The pie charts below summarize the lifetime distribution (<5 years, 5-8 years, lasting the entire or exceeding the study period).

### 3.3. Offshore oil and gas platforms in 2025

In the first quarter of 2025, offshore oil and gas infrastructure was clearly concentrated in the Persian Gulf (1,731 platforms) and the Gulf of Mexico (1,641), while the North Sea had significantly fewer platforms with 356 installations (Figure 9).

Clear spatial concentrations can be seen within the regions. In the Persian Gulf, most installations were located in the southwestern half of the basin along the coasts of Saudi Arabia (802), Qatar (196), and the United Arab Emirates (523), while Iran had comparatively fewer platforms (160) despite its large exclusive economic zone, as Iranian oil and gas production is predominantly located onshore [58]. In the Gulf of Mexico, the platforms were entirely located in the exclusive economic zones of United States and Mexico, with the US dominating by far (1,386 versus 255 platforms). In the North Sea, offshore infrastructure was spread across several countries, led by the United Kingdom with 178 installations, followed by EU member states with a total of 137 installations and Norway with 41 platforms. Offshore activities in the three regions were thus dominated by a few key players, led by the United States, followed by Saudi Arabia and the United Arab Emirates.

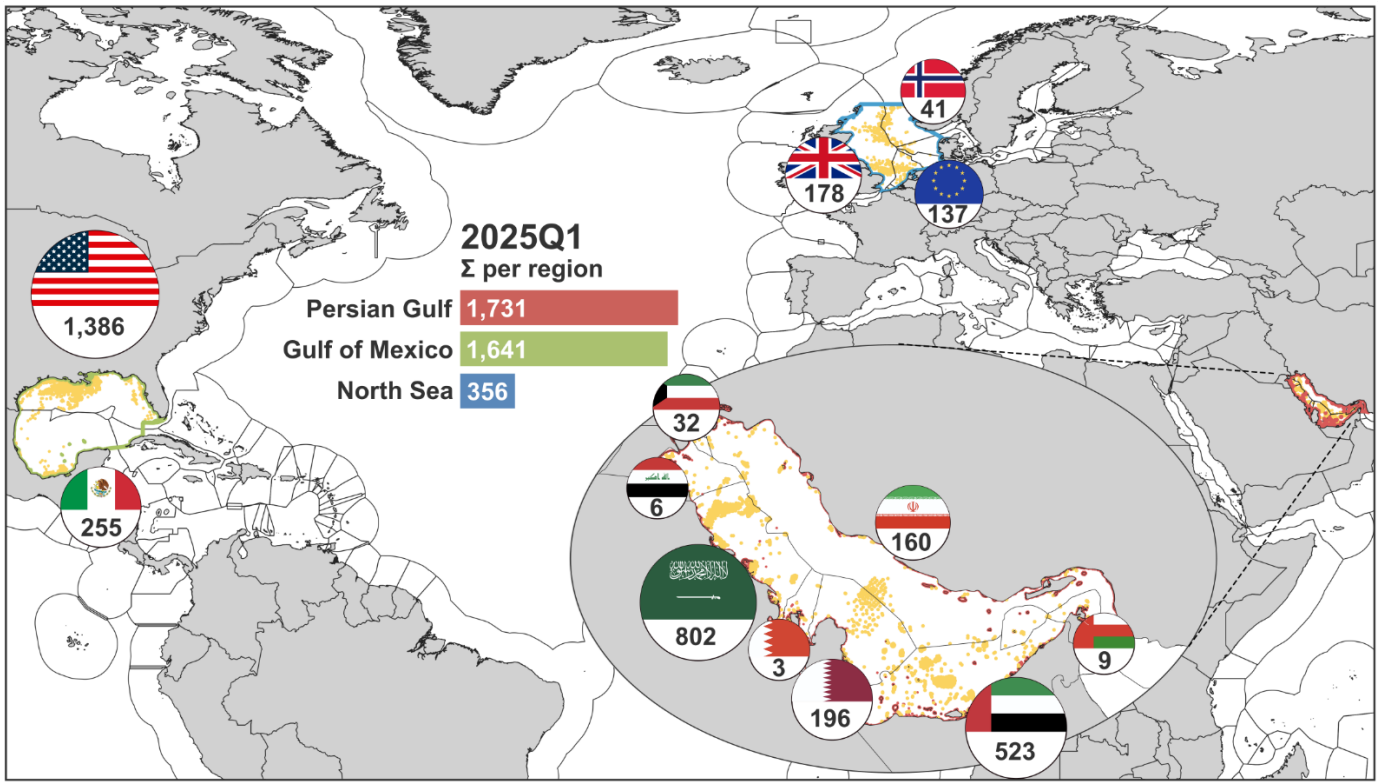

**Figure 9.** Number and distribution of offshore oil and gas platforms in the Gulf of Mexico, the North Sea, and the Persian Gulf (enlarged section) in the first quarter of 2025. In addition, the platform counts are shown by affiliation to exclusive economic zones (EU member states in the North Sea are aggregated). The three most important contributors in terms of platform counts in the offshore oil and gas sector of these regions are the United States, followed by Saudi Arabia and the United Arab Emirates.

The spatial distribution of offshore oil and gas infrastructure and the temporal and spatial dynamics of new installations, decommissioning, and relocations are shown in Figure 10. At the same time, Sentinel-1 images demonstrate the structural diversity of offshore infrastructure. In the Gulf of Mexico, offshore infrastructure was heavily concentrated in the northern part of the basin along the Texas-Louisiana Shelf. Particularly high platform densities occurred off the coast of Louisiana (hotspot E), and along the Texas coast in the northwest of the Gulf. Mexico's offshore oil and gas infrastructure was mainly concentrated in the southern part of the basin in the Bay of Campeche off the Mexican coast (hotspot F). The color coding shows that new installations continue to be added in this region. In the North Sea, offshore infrastructure was concentrated primarily in the central basin between the UK and Norway and in the southwestern part along the Dutch and British continental shelves (hotspot G). Compared to other regions, the North Sea had a higher number of decommissioned or relocated platforms (gray dots). At the same time, there were numerous newer installations. This pattern reflects the previously observed dynamics, which are characterized by both the decommissioning and the use of temporary or mobile offshore units such as jack-up rigs or drillships. Platform density was particularly high in the Persian Gulf in the southwestern half of the basin along the coasts of Saudi Arabia, Qatar, and the United Arab Emirates. The detailed views show large-scale platform clusters, particularly off the Saudi Arabian coast in the northernmost part of the Saudi EEZ (hotspot H). The color distribution indicates that a significant portion of this infrastructure has been expanded in recent years, confirming the previously identified increase in installations in Saudi Arabia and throughout the entire region.

## 4. Discussion

### 4.1. Implications and interpretation of results

This study demonstrates that freely available Earth observation data, combined with deep learning-based object detection, enable high-resolution, time series analysis of offshore oil and gas infrastructure. Based on the Sentinel-1 archive and the user-friendly YOLO detection framework, a consistent time series of offshore platforms between 2017 and 2025 was created. This not only allowed platform locations to be recorded, but also enabled additional spatial-temporal information such as size, distance to coast, water depth, affiliation to exclusive economic zones, and installation and removal data to be derived.

The results reveal significant regional differences in the development of offshore oil and gas infrastructure between 2017 and 2025, reflecting varying dynamics in the offshore oil and gas sector. While the Persian Gulf has continued to expand its offshore structures until mid-2024, the Gulf of

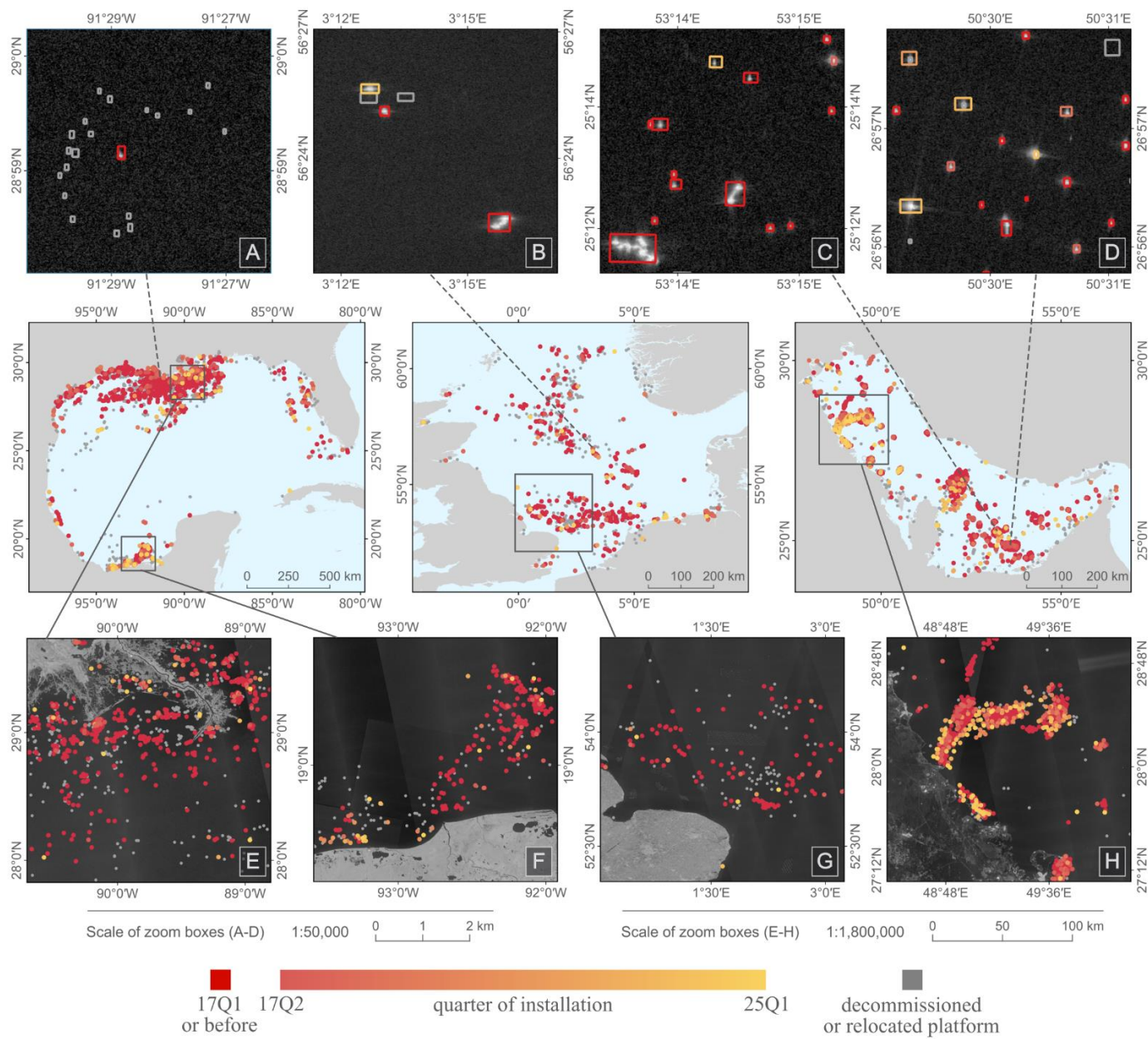

**Figure 10.** Map of the three study regions with detailed zoom views on hotspot oil and gas platform regions showing the Sentinel-1-derived active offshore oil and gas platform status from 2025Q1. Colored points indicate the quarter of installation (gradient from 2017Q1 or earlier to 2025Q1). Gray points mark decommissioned or relocated platforms.

Mexico and, in particular, the North Sea showed a decline since peaking around 2018-2020. In the North Sea, this trend is related to energy policy transformation processes in the context of long-term net-zero strategies, while in the Gulf of Mexico it is primarily related to the increasing decommissioning of older platforms, with total production staying high or rising. The spatial characteristics of platform locations, such as platform size, distance from the coast, or water depth, showed largely stable patterns with no noticeable trends over the entire study period. Changes in the offshore sector are therefore reflected less in temporal shifts and more in region-specific structures, such as greater offshore expansion and platform size in the North Sea and more clustered platforms and more large platform complexes in the Persian Gulf. The analysis of platform lifetime also indicated structural changes in the type of infrastructure used. While the majority of structures continued to exist for very long periods (beyond our study period of over 8 years), there is an increasing presence of short-lived platforms in all three regions (especially in the North Sea). This points to the growing importance of temporary and mobile offshore units such as jack-up platforms or drillships and reflects an increasing flexibility in offshore production strategies. Technological advances and more efficient extraction methods are increasingly making it possible to achieve high production levels with fewer platforms, meaning that changes in the offshore sector are reflected less in the number of platforms and more in their technical specification and flexibility of use. The approach developed in this study for determining installation and removal data for platform locations adds an important temporal dimension to existing platform datasets and, for the first time, allows such changes in the persistence and mobility of offshore infrastructure to be systematically monitored and quantified.

### 4.2. Broader implications of the dataset and future perspectives

Beyond merely recording infrastructure, the dataset thus provides a basis for analyzing developments in the offshore oil and gas sector in a more differentiated manner and placing them in a broader context. The quarterly time interval allows potential correlations between infrastructural changes and external influences to be examined, such as energy policy decisions, regulatory measures, new uses or reuses for former platform sites, for example as $CO_2$ storage sites or artificial reefs. A spatial expansion of the dataset to global scales and greater contextualization through expertise from related disciplines could provide additional insights into the dynamics of the offshore oil and gas sector in the future. This is also relevant because the maritime space as a whole is undergoing a phase of intensive and increasing use. In addition to offshore oil and gas production, offshore wind energy, shipping, fishing, aquaculture, and marine conservation also compete for marine areas, leading to increased conflicts of use and planning requirements. This competition for space is particularly evident in heavily used regions such as the North Sea. The additional information derived from this study, such as water depth, distance from the coast, platform size, and installation and removal data, can therefore help to map marine infrastructure and its development more precisely and thus provide a more robust basis for monitoring, spatial planning, and accompanying research in a phase of intensive anthropogenic use of the oceans. At the same time, such spatially and temporally consistent datasets also enable improved investigation of the potential impacts of offshore infrastructure on marine ecosystems. The study demonstrates the value that freely available Earth observation data can have for independent research in a dynamic and industry-driven sector.

### 4.3. Limitations of our study

The analysis is based exclusively on the Sentinel-1 acquisitions that are effectively available. These do not provide completely homogeneous coverage of the global oceans, as the acquisition schedule follows mission-specific priorities. Although Sentinel-1 focuses on coastal regions and areas with high maritime activity, thereby covering most industrial offshore activities, part of the central basin in the Gulf of Mexico (approx. 22-28°N, 84-88°W) is not consistently covered. Furthermore, the detection of very small platforms is limited by the spatial resolution of Sentinel-1 data (10 m). Structures that are close to or smaller than the pixel size can be difficult to distinguish from the sea surface, especially in rough sea conditions or in coastal areas. This can lead to a slight underestimation of very small platforms, especially in regions with high platform density such as the Gulf of Mexico. Further misclassifications may arise due to noise artifacts or other metallic objects at sea. These include, for example, buoys, lighthouses, or rocks. To reduce noise, a pixel value threshold was integrated in post-processing. Additional manual filtering or removal of misclassifications was intentionally not performed in order to ensure the reproducibility and scalability of the automated workflow.

The derivation of platform lifetime was based on the first and last detection of a platform at a location. Some platforms could not be detected in every single quarter. Therefore, temporal gaps were filled in when platforms were detected at previous and subsequent points in time in order to derive a coherent platform lifespan. Detections that occurred in only a single quarter were purposely retained so as not to exclude potential short-term offshore activities. More rigorous filtering of such isolated events could further refine the dataset in future work.

In addition, the integration of additional SAR missions, such as future Sentinel-1 satellites, will further improve spatial and temporal coverage and enable broader application of the method, allowing the dataset to be extended to other regions and to a global analysis in the future. The fusion of remote sensing data from multiple sources could also help to further increase the completeness and accuracy of the platform dataset, for example, to close the acquisition gap in the Gulf of Mexico or to improve the detection of very small platform structures. In addition, future analyses could incorporate other sensor systems, such as those in the near-infrared (NIR) range, to derive additional information such as the activity status of platforms and to better qualify and contextualize infrastructural developments after decommissioning through possible reuse, for example as $CO_2$ storage or artificial reefs.

## 5. Conclusion

This study proposed an approach for the spatiotemporal analysis of recent developments in the offshore oil and gas sector in three of the world's most important production regions: the Gulf of Mexico, the Persian Gulf, and the North Sea. The entire workflow was based on freely available Earth observation data, the Sentinel-1 archive, supplementary GIS datasets, and the recently developed and validated object detection model for the automated detection of offshore oil and gas platforms by Spanier et al. 2026 [42]. On this basis, a consistent, high-resolution, quarterly time series of platform locations was created for the period 2017 to 2025. Accurate location information and attributes relating to distribution, size, affiliation to exclusive economic zones, and lifespan (installation and removal dates) were derived and combined into a vectorized dataset. A total of 3,728 platforms were identified for 2025, including 1,731 in the Persian Gulf, 1,641 in the Gulf of Mexico, and 356 in the North Sea. The United States had the highest number with 1,386 platforms, followed by Saudi Arabia with 802 and the United Arab Emirates with 523 platforms. Between 2017 and 2025, a significant dynamic in offshore oil and gas infrastructure was observed. A total of 2,763 platforms were newly constructed or relocated to new sites, while 2,718 platforms were decommissioned or relocated. This reflects an observed structural shift from long-term installed platforms to mobile and floating production units, a trend particularly evident in the North Sea. While an increasing number of platforms were decommissioned in the Gulf of Mexico, the Persian Gulf showed a significant upward trend in platform counts until 2024, while platforms in the North Sea declined, which is consistent with the energy policy transformation process. Around 90% of all platforms were located in water depths of less than 100 m, while the deepest identified infrastructure, the Turritella FPSO in the Stones field in the Gulf of Mexico, operated at a depth exceeding 2,900 m. Spatially, there are highly clustered platform distributions in the Gulf of Mexico and the Persian Gulf, while the platforms in the North Sea were located further offshore and were more dispersed. Given the growing number of artificial infrastructure in marine ecosystems, these findings underscore the need for integrated planning approaches in coastal and marine areas. The developed dataset and associated attribute information make an important contribution to supporting offshore monitoring, and assessing and mitigating potential environmental impacts. At the same time, the approach presented enables the timely and automated recording of offshore infrastructure and supplements existing platform information from government or industry sources. Since the entire workflow is based on freely accessible data and tools, it is spatially and temporally scalable and offers potential for applications in other regions or for global analyses.

## Acknowledgment

The authors gratefully acknowledge the computational and data resources provided through the joint high-performance data analytics (HPDA) project "terrabyte" of the German Aerospace Center (DLR) and the Leibniz Supercomputing Center (LRZ). The authors gratefully acknowledge ESA's Copernicus program for providing free access to the Sentinel-1 data and the Google Earth Engine platform for preprocessing and making the data accessible.

## Data avalability statement

The Offshore Platform Dataset (OPD v1.0.0) that represents the findings of this study is openly available via Zenodo: Spanier, R., Hoeser, T., Truckenbrodt, J., Bachofer, F., Kuenzer, C. (2026). *OPD v1.0.0: Sentinel-1 derived dataset of offshore oil and gas platform locations, lifespans, and spatial attributes.* [Dataset]. Zenodo. https://doi.org/10.5281/zenodo.19046824.

## Author contribution

**Robin Spanier**: Conceptualisation, methodology, code development, data curation, validation, visualisation, original manuscript writing. **Thorsten Hoeser**: Conceptualization, data curation, manuscript reviewing. **John Truckenbrodt**: data curation, manuscript reviewing. **Felix Bachofer**: Supervision, manuscript reviewing. **Claudia Kuenzer**: Conceptualization, supervision, manuscript reviewing.

## Funding

This research received no external funding.

## Disclosure statement

The authors report no conflict of interest.

## References

[1] Y. Liu, C. Hu, Y. Dong, B. Xu, W. Zhan, and C. Sun, "Geometric accuracy of remote sensing images over oceans: The use of global offshore platforms," *Remote Sensing of Environment*, vol. 222, pp. 244–266, 2019, doi: 10.1016/j.rse.2019.01.002.

[2] A. B. Bugnot *et al.*, "Current and projected global extent of marine built structures," *Nat Sustain*, vol. 4, no. 1, pp. 33–41, 2021, doi: 10.1038/s41893-020-00595-1.

[3] J. Wang, M. Li, Y. Liu, H. Zhang, W. Zou, and L. Cheng, "Safety assessment of shipping routes in the South China Sea based on the fuzzy analytic hierarchy process," *Safety Science*, vol. 62, pp. 46–57, 2014, doi: 10.1016/j.ssci.2013.08.002.

[4] B. J. Williamson, P. Blondel, E. Armstrong, P. S. Bell, C. Hall, and J. J. Waggitt, "A Self-Contained Subsea Platform for Acoustic Monitoring of the Environment Around Marine Renewable Energy Devices–Field Deployments at Wave and Tidal Energy Sites in Orkney, Scotland," *IEEE J. Oceanic Eng.*, vol. 41, no. 1, pp. 67–81, 2016, doi: 10.1109/JOE.2015.2410851.

[5] P. E. Posen, K. Hyder, M. Teixeira Alves, N. G. Taylor, and C. P. Lynam, "Evaluating differences in marine spatial data resolution and robustness: A North Sea case study," *Ocean & Coastal Management*, vol. 192, p. 105206, 2020, doi: 10.1016/j.ocecoaman.2020.105206.

[6] P. Ma, M. Macdonald, S. Rouse, and J. Ren, "Automatic Geolocation and Measuring of Offshore Energy Infrastructure With Multimodal Satellite Data," *IEEE J. Oceanic Eng.*, vol. 49, no. 1, pp. 66–79, 2024, doi: 10.1109/JOE.2023.3319741.

[7] X. Zhang, M. Qiu, S. Tao, X. Ge, and M. Wang, "OPDNet: An Offshore Platform Detection Network Based on Bitemporal Bimodal Remote-Sensing Images and a Pseudo-Siamese Structure," *IEEE J. Sel. Top. Appl. Earth Observations Remote Sensing*, vol. 18, pp. 6409–6421, 2025, doi: 10.1109/JSTARS.2025.3539674.

[8] K. Sadeghi, "An overview of design, analysis, construction and installation of offshore petroleum platforms suitable for Cyprus oil/gas fields," *GAU J. Soc. Appl. Sci*, vol. 2, no. 4, pp. 1–16, 2007.

[9] OECD, *The Ocean Economy to 2050*.

[10] OPEC, "World Oil Outlook 2050," Vienna, 2025. [Online]. Available: https://www.opec.org/assets/assetdb/woo-2025.pdf

[11] R. W. K. Potter and B. C. Pearson, "Assessing the global ocean science community: understanding international collaboration, concerns and the current state of ocean basin research," *npj Ocean Sustain*, vol. 2, no. 1, 2023, doi: 10.1038/s44183-023-00020-y.

[12] P. C. J. Da Vidal *et al.*, "Decommissioning of offshore oil and gas platforms: A systematic literature review of

factors involved in the process," *Ocean Engineering*, vol. 255, p. 111428, 2022, doi: 10.1016/j.oceaneng.2022.111428.

[13] A. R. Gates and D. O. B. Jones, "Ecological role of offshore structures," *Nat Sustain*, vol. 7, no. 4, pp. 383–384, 2024, doi: 10.1038/s41893-024-01316-8.

[14] R. Spanier and C. Kuenzer, "Marine Infrastructure Detection with Satellite Data—A Review," *Remote Sensing*, vol. 16, no. 10, p. 1675, 2024, doi: 10.3390/rs16101675.

[15] D. Petrobras, *Oil platform P-51 (Brazil).jpg*. [Online]. Available: https://commons.wikimedia.org/w/index.php?curid=5621984

[16] A. Murray, *Visit the North Sea oil field used to store greenhouse gas*: BBC, 2026. [Online]. Available: https://www.bbc.com/news/articles/cq5y7dd284do

[17] N. Skopljak, *Denmark grants its first-ever approval for CO2 storage facility*: Offshore Energy, 2025. [Online]. Available: https://www.offshore-energy.biz/denmark-grants-its-first-ever-approval-for-co2-storage-facility/

[18] S. Y. Toh, C. MacBeth, J. Landa, and H. Heidari, "Exploring 4D seismic potential for monitoring CO2 injection in depleted North Sea gas fields," *International Journal of Greenhouse Gas Control*, vol. 148, p. 104529, 2025, doi: 10.1016/j.ijggc.2025.104529.

[19] A. P. Singh et al., "Monitoring long-term storage of CO2 in a gas and condensate field in the North Sea off the coast of Norway using seismic methods," *Geophysics*, vol. 90, no. 4, M167-M180, 2025, doi: 10.1190/geo2024-0715.1.

[20] OECD, "OECD Science, Technology and Industry Working Papers," 2021.

[21] D. March, K. Metcalfe, J. Tintoré, and B. J. Godley, "Tracking the global reduction of marine traffic during the COVID-19 pandemic," *Nature communications*, vol. 12, no. 1, p. 2415, 2021, doi: 10.1038/s41467-021-22423-6.

[22] European Commission and Directorate-General for Environment, *Environmental impact assessment of projects – Rulings of the Court of Justice of the European union*: Publications Office of the European Union, 2022.

[23] J. Virdin et al., "The Ocean 100: Transnational corporations in the ocean economy," *Science advances*, vol. 7, no. 3, 2021, doi: 10.1126/sciadv.abc8041.

[24] B. A. Wong, C. Thomas, and P. Halpin, "Automating offshore infrastructure extractions using synthetic aperture radar & Google Earth Engine," *Remote Sensing of Environment*, vol. 233, p. 111412, 2019, doi: 10.1016/j.rse.2019.111412.

[25] C. Sun, Y. Liu, S. Zhao, and S. Jin, "Estimating offshore oil production using DMSP-OLS annual composites," *ISPRS Journal of Photogrammetry and Remote Sensing*, vol. 165, pp. 152–171, 2020, doi: 10.1016/j.isprsjprs.2020.05.019.

[26] Y. Liu, C. Sun, Y. Yang, M. Zhou, W. Zhan, and W. Cheng, "Automatic extraction of offshore platforms using time-series Landsat-8 Operational Land Imager data," *Remote Sensing of Environment*, vol. 175, pp. 73–91, 2016, doi: 10.1016/j.rse.2015.12.047.

[27] C. A. Baumhoer, A. J. Dietz, K. Heidler, and C. Kuenzer, "IceLines - A new data set of Antarctic ice shelf front positions," *Scientific data*, vol. 10, no. 1, p. 138, 2023, doi: 10.1038/s41597-023-02045-x.

[28] T. Esch et al., "World Settlement Footprint 3D - A first three-dimensional survey of the global building stock," *Remote Sensing of Environment*, vol. 270, p. 112877, 2022, doi: 10.1016/j.rse.2021.112877.

[29] T. Hoeser and C. Kuenzer, "Global dynamics of the offshore wind energy sector monitored with Sentinel-1: Turbine count, installed capacity and site specifications," *International Journal of Applied Earth Observation and Geoinformation*, vol. 112, p. 102957, 2022, doi: 10.1016/j.jag.2022.102957.

[30] UNCTAD, *Exploring space technologies for sustainable development*. [Online]. Available: https://unctad.org/publication/exploring-space-technologies-sustainable-development

[31] J. Yang et al., "The role of satellite remote sensing in climate change studies," *Nature Clim Change*, vol. 3, no. 10, pp. 875–883, 2013, doi: 10.1038/nclimate1908.

[32] L. Cheng, K. Yang, L. Tong, Y. Liu, and M. Li, "Invariant triangle-based stationary oil platform detection from multitemporal synthetic aperture radar data," *J. Appl. Remote Sens*, vol. 7, no. 1, p. 73537, 2013, doi: 10.1117/1.JRS.7.073537.

[33] A. Marino, D. Velotto, and F. Nunziata, "Offshore Metallic Platforms Observation Using Dual-Polarimetric TS-X/TD-X Satellite Imagery: A Case Study in the Gulf of Mexico," *IEEE J. Sel. Top. Appl. Earth Observations Remote Sensing*, vol. 10, no. 10, pp. 4376–4386, 2017, doi: 10.1109/JSTARS.2017.2718584.

[34] J. Zhang, Q. Wang, and F. Su, "Automatic Extraction of Offshore Platforms in Single SAR Images Based on a Dual-Step-Modified Model," *Sensors (Basel, Switzerland)*, vol. 19, no. 2, 2019, doi: 10.3390/s19020231.

[35] W. Xu et al., "Proliferation of offshore wind farms in the North Sea and surrounding waters revealed by satellite image time series," *Renewable and Sustainable Energy Reviews*, vol. 133, p. 110167, 2020, doi: 10.1016/j.rser.2020.110167.

[36] H. Zhu, G. Jia, Q. Zhang, S. Zhang, X. Lin, and Y. Shuai, "Detecting Offshore Drilling Rigs with Multitemporal NDWI: A Case Study in the Caspian Sea," *Remote Sensing*, vol. 13, no. 8, p. 1576, 2021, doi: 10.3390/rs13081576.

[37] C. R. Jackson and J. R. Apel, "Synthetic aperture radar: marine user's manual," 2004.

[38] T. Hoeser, S. Feuerstein, and C. Kuenzer, "DeepOWT: a global offshore wind turbine data set derived with deep learning from Sentinel-1 data," *Earth Syst. Sci. Data*, vol. 14, no. 9, pp. 4251–4270, 2022, doi: 10.5194/essd-14-4251-2022.

[39] F. S. Paolo et al., "Satellite mapping reveals extensive industrial activity at sea," *Nature*, vol. 625, no. 7993, pp. 85–91, 2024, doi: 10.1038/s41586-023-06825-8.

[40] T. Zhang, B. Tian, D. Sengupta, L. Zhang, and Y. Si, "Global offshore wind turbine dataset," *Scientific data*,

vol. 8, no. 1, p. 191, 2021, doi: 10.1038/s41597-021-00982-z.

[41] T. Hoeser and C. Kuenzer, "SyntEO: Synthetic dataset generation for earth observation and deep learning – Demonstrated for offshore wind farm detection," *ISPRS Journal of Photogrammetry and Remote Sensing*, vol. 189, pp. 163–184, 2022, doi: 10.1016/j.isprsjprs.2022.04.029.

[42] R. Spanier, T. Hoeser, and C. Kuenzer, "Deep learning-based object detection of offshore platforms on Sentinel-1 imagery and the impact of synthetic training data," *International Journal of Remote Sensing*, vol. 47, no. 5, pp. 1–25, 2026, doi: 10.1080/01431161.2026.2612908.

[43] Flanders Marine Institute, "IHO Sea Areas, version 3," 2018.

[44] J.-P. Ducrotoy, M. Elliott, and V. N. de Jonge, "The North Sea," *Marine pollution bulletin*, vol. 41, 1-6, pp. 5–23, 2000, doi: 10.1016/S0025-326X(00)00099-0.

[45] J. Sündermann and T. Pohlmann, "A brief analysis of North Sea physics," *Oceanologia*, vol. 53, no. 3, pp. 663–689, 2011, doi: 10.5697/oc.53-3.663.

[46] IEA, "Oil 2025: Analysis and forecast to 2030," 2025. [Online]. Available: https://iea.blob.core.windows.net/assets/c0087308-f434-4284-b5bb-bfaf745c81c3/Oil2025.pdf

[47] V. Prescott and C. Schofield, "The Persian Gulf," in *The Maritime Political Boundaries of the World*, V. Prescott and C. Schofield, Eds.: Brill Nijhoff, 2004, pp. 497–518.

[48] J. Kämpf and M. Sadrinasab, "The circulation of the Persian Gulf: a numerical study," *Ocean Sci.*, vol. 2, no. 1, pp. 27–41, 2006, doi: 10.5194/os-2-27-2006.

[49] EIA, *Amid regional conflict, the Strait of Hormuz remains critical oil chokepoint.* [Online]. Available: https://www.eia.gov/todayinenergy/detail.php?id=65504

[50] L. McKinney *et al.,* "The Gulf of Mexico: An Overview," *Oceanog*, vol. 34, no. 1, pp. 30–43, 2021, doi: 10.5670/oceanog.2021.115.

[51] K. El-Darymli, P. McGuire, D. Power, and C. Moloney, "Target detection in synthetic aperture radar imagery: a state-of-the-art survey," *J. Appl. Remote Sens*, vol. 7, no. 1, p. 71598, 2013, doi: 10.1117/1.JRS.7.071598.

[52] E. S. A. ESA, *Sentinel-1*. [Online]. Available: https://sentiwiki.copernicus.eu/web/sentinel-1

[53] GEE and ESA, *Sentinel-1 SAR GRD Collection (COPERNICUS/S1\_GRD).* [Online]. Available: https://developers.google.com/earth-engine/datasets/catalog/COPERNICUS_S1_GRD

[54] A. Wang *et al.,* "YOLOv10: Real-Time End-to-End Object Detection," 2024.

[55] EMODnet, *EMODnet Human Activities, Energy, Wind Farms (Dataset).* [Online]. Available: https://emodnet.ec.europa.eu/geonetwork/srv/eng/catalog.search#/metadata/8201070b-4b0b-4d54-8910-abcea5dce57f

[56] GEBCO, *GEBCO Bathymetry 2024 dataset.* [Online]. Available: https://download.gebco.net/#

[57] L. Si, S. Zhou, I. Irakulis-Loitxate, J. Roger, and L. Guanter, "The Offshore Oil and Gas Platforms (OOGPs) dataset based on satellite data spanning 2017 to 2023," 2026.

[58] EIA, "Country Analysis Brief: Iran," Washington DC, 2024. [Online]. Available: https://www.eia.gov/international/content/analysis/countries_long/Iran/pdf/Iran%20CAB%202024.pdf